# Optical Photon Reassignment Microscopy (OPRA)

Stephan Roth[1,2], Colin J. R. Sheppard[3], Kai Wicker[1,2] and Rainer Heintzmann[1,2,4]

[1]Institute of Photonics Technology, Albert-Einstein-Str.9, 07745 Jena, Germany

[2]Institute of Physical Chemistry, Abbe Center of Photonics, Friedrich-Schiller-University Jena, Helmholtzweg 4, 07743 Jena, Germany

[3]Nanophysics, Instituto Italiano di Tecnologia, via Morego 30, 16163 Genova, Italy

[4]King's College London, Randall Division, NHH, SE1 1UL, London, U.K.

*Corresponding author: heintzmann@gmail.com

ABSTRACT

To enhance the resolution of a confocal laser scanning microscope the additional information of a pinhole plane image taken at every excitation scan position can be used [1]. This photon reassignment principle is based on the fact that the most probable position of an emitter is at half way between the nominal focus of the excitation laser and the position corresponding to the (off centre) detection position. Therefore, by reassigning the detected photons to this place, an image with enhanced detection efficiency and resolution is obtained. Here we present optical photon reassignment microscopy (OPRA) which realises this concept in an all-optical way obviating the need for image-processing. With the help of an additional intermediate optical beam expansion between descanning and a further rescanning of the detected light, an image with the advantages of photon reassignment can be acquired. Due to its simplicity and flexibility this method has the potential to enhance the performance of nearly every laser scanning microscope and is therefore expected to play an important role in future systems.

Abbrevations:

| | | |
|---|---|---|
| | CLSM | confocal laser scanning microscope |
| | ISM | images scanning microscopy |
| | LSM | laser scanning microscope |
| | OPRA | optical photon reassignment |
| | PSF | point spread function |
| | RESOLFT | reversible saturable optical fluorescence transitions |
| | sCMOS | scientific complementary metal–oxide–semiconductor |
| | STED | stimulated emission depletion |

INTRUDUCTION

Confocal laser scanning microscopy (CLSM) is an established tool in fluorescence microscopy and well-known for its optical sectioning ability and high contrast [2,3]. These characteristics are achieved by using detectors with a high dynamic range and collecting the emitted light through a pinhole, which is usually aligned to the position of the excitation focus (thus the name "confocal"). The resulting image is constructed by assigning the detected intensity to the corresponding excitation scan position. In 1982 it was shown that it is possible to achieve enhanced resolution by applying an off-axis pinhole [4]. In 1988 pinhole plane image detection and computational reassignment to a position half way between nominal excitation and detection position was proposed [1], to improve detection efficiency and resolution. Note that for identical excitation and emission point spread functions (PSF) this reassigned position corresponds to the most probable position of an emitter in the sample. Recent work applied this principle in single [5] and multispot excitation [6] to the imaging of biological samples.

Here we present optical photon reassignment microscopy (OPRA). It is an optical realisation of these computer based methods which avoids the need for data processing. Furthermore at a different scaling ratio, our method is applicable to the direct visualisation of high-resolution imaging methods like STED.

BACKGROUND

In normal CLSM the detected intensity values of every scanning position are recorded with an integrating detector such as a photo-multiplier-tube (PMT) or an avalanche photo-diode. If a whole image of the pinhole plane is recorded at each scan position the acquired data-set of a single focal slice can be viewed as a 4 dimensional set of data (intensity values in dependency of xy scan and xy pinhole-plane image-position). Such data can be processed in several ways, for example allowing for a retroactive choice of the pinhole diameter and/or applying multi-view deconvolution [8]. Photon reassignment microscopy [1,5] is based on the insight that the most probable origin of the detected photons is at maximum of the joint probability function (i.e. the product of the individual probability functions) of excitation and (off-centre) detection. This is contrary to a CLSM where all detected photons are assigned to the nominal excitation position *s*.

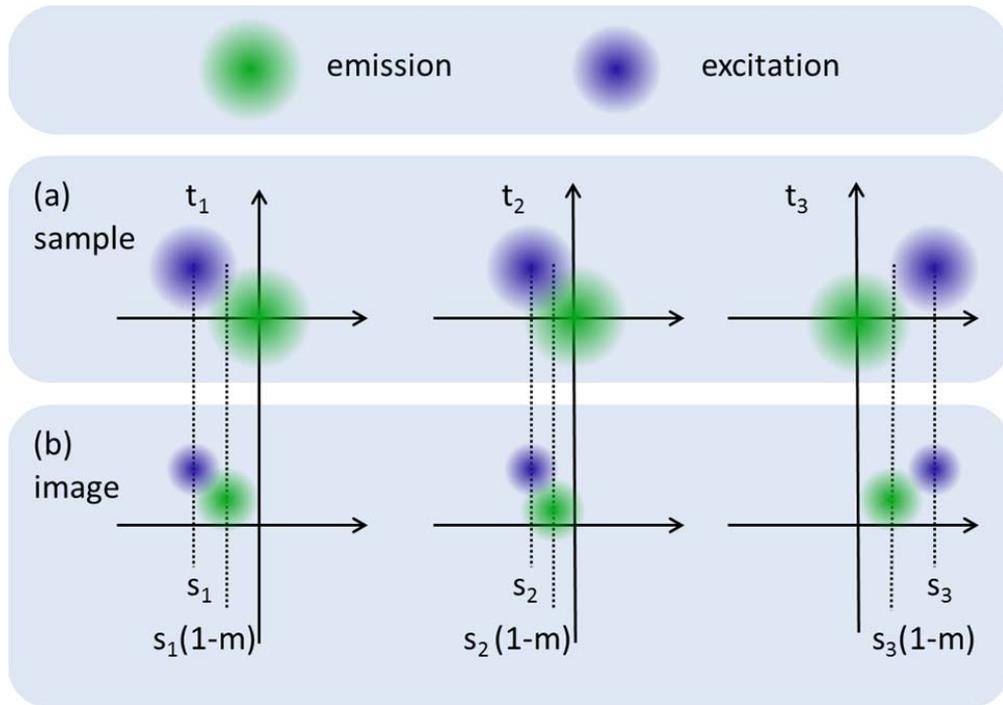

**Figure 1 The principle of OPRA.** The figure shows the imaging process of one point source at different times. The fluorophore is placed at the point of origin in the sample plane (a). The emitted photons are imaged to different positions in the image plane (b) according to the excitation positions *s*. If the general magnification of the microscope is neglected and the intermediate magnification is *m*=0.5 the photons are reassigned to half the distance between the nominal excitation position *s* and the position of the detected photon without intermediate magnification. In normal scanning microscopy the photons are always assigned to position *s*. Note that the brightness changes of the green emitter caused by the variation in excitation are not shown in this scheme.

In OPRA a similar reassignment to the optimal emission location is achieved by an intermediate beam expansion between descanning and a subsequent rescanning. This is illustrated in Fig. 1 at 3 successive time points. The upper row (1a) depicts the situation in the sample, where a scanned excitation beam (blue) together with a fixed emitter (green) at the origin is shown. The lower row (1b) refers to the final image plane. Due to the intermediate beam expansion, the emission PSF is reduced in size. As a further consequence the image of the emitter is now found at *s*(1-*m*), with *s* being the nominal image position of the centre of the excitation focus and *m* being the intermediate magnification. At non-uniform intermediate magnification (*m*≠1) the image of the emitter now performs a small scan on the final image plane, changing its brightness under the influence of the excitation spot.

We now aim to find an analytical description of the PSF of the overall system. The emission PSF has also undergone the intermediate beam expansion *m* and can therefore be written as $h(x/m)$, positioned at $s(1-m)$. Thus, the current image generated by the point source at the point of origin is $h_{em}\left(\frac{x-s(1-m)}{m}\right)$, where *s* is the nominal scan position and x the image coordinate (measured in sample coordinates).

The total image of our point emitter is formed by integrating over all scan positions *s*:

$$h_{total}(x) = \int h_{ex}(-s)\, h_{em}\left(\frac{x-s(1-m)}{m}\right) ds. \qquad (1)$$

This can be written using the convolution operator $\otimes$ as

$$h_{total}(x) = [h'_{ex} \otimes h'_{em}]\left(\frac{x}{1-m}\right), \tag{2}$$

where $h'_{em}(x) := h_{em}\left(\frac{x*(1-m)}{m}\right)$ and the symmetrical excitation PSF $h'_{ex}(x) := h_{ex}(-x)$ are used.

If we assume a Gaussian shaped excitation and emission function $f(x) = \exp\left(-\frac{1}{2}\left(\frac{x}{\sigma}\right)^2\right)$ (standard deviation $\sigma_{ex}$ for the excitation and corresponding $\sigma_{em}$ for the emission function) the integral can be solved analytically. The final PSF is found to have the standard deviation

$$\sigma_{OPRA}{}^2 = \sigma_{em}^2 m^2 + \sigma_{ex}^2 (1-m)^2. \tag{3}$$

The minimal total extent is found as:

$$m = \frac{\sigma_{ex}{}^2}{\sigma_{ex}{}^2 + \sigma_{em}{}^2}. \tag{4}$$

Thus the additional expansion *m* should be adjusted to the different width of the excitation and emission PSFs. This difference can be induced by the Stokes shift of the used fluorophores or is a feature of the microscopy technique itself, to which OPRA is applied (e.g. STED microscopy, where σ<sub>ex</sub> is significantly smaller than σ<sub>em</sub>).

If we assume a beam expansion value of *m*=0.5 (which makes a wider beam and therefore produces a smaller spot when focussed) we get a rough estimate for the resolution ability of OPRA

$$\sigma_{OPRA} = 0.5 \cdot \sqrt{\sigma_{ex}^2 + \sigma_{em}^2}. \tag{5}$$

If no Stokes shift is considered ($\sigma_{ex} = \sigma_{em}$) we obtain a resolution improvement of $\sqrt{2}$ over what we would expect for confocal detection with a closed pinhole. This shows that reassignment microscopy realizes high resolution at the theoretical detection efficiency of a widefield microscope. OPRA attains the same characteristics as computational reassignment without the need for high-speed pinhole cameras and without the increased read-noise of multiple readouts. This raises the acquisition speed as the whole image is acquired in only one exposure frame. An additional pinhole can also be integrated in OPRA (before rescanning) to achieve confocal sectioning.

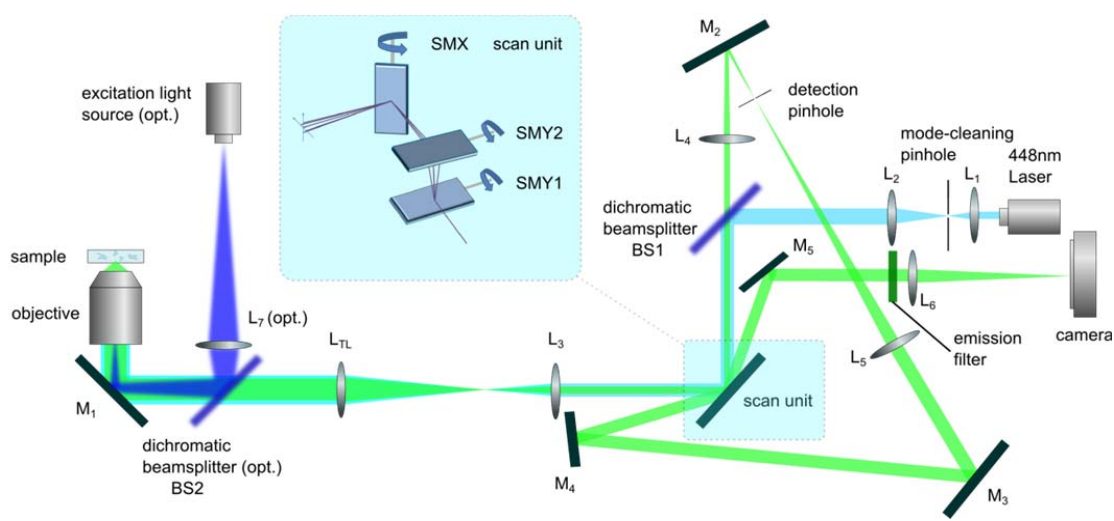

**Figure 2 OPRA Setup**. The 488nm laser passes a clean-up (lens $L_1$ and $L_2$ with appending focal length $f_1$=50mm; $f_2$=100mm) and is directed to a dichromatic beam splitter (BS1, detection wavelength bigger than 505nm). The scan unit in detail is shown in the inset. Here two y-scanning-mirrors (SMY) are used to project the spot to the rotation axes of the x-scanning-mirror (SMX, 15 kHz). After the scanning unit the beam passes a second beam expander and is directed to the objective. The returning fluorescent light is descanned and separated from the excitation light using the dichromatic beam splitter (BS1). After descanning the fluorescent beam is expanded by a factor of two ($f_4$=200mm; $f_5$=400mm). The adjustable detection pinhole between the lenses $L_4$ and $L_5$ can be used to achieve confocal sectioning (not in measurements). After the expansion the beam is rescanned using the same scanning system and projected via the lens $L_6$ ($f_6$=200mm) to the camera. To compare the images with a widefield setup an excitation light source, the optional (opt.) lenses $L_7$, and a dichromatic beam splitter (BS2) were added in this configuration, while the scanner does not move and the detection pinhole is removed.

METHODS

Figure 2 shows the experimental setup. The illumination part is a normal laser scanning setup, which creates a moving illumination spot in the sample plane. The beam of the excitation laser (Coherent, Sapphire LP 488nm) is sent through a beam expander ($L_1$ and $L_2$) to a dichromatic beam splitter (BS1, AHF Analysetechnik Tübingen, ZT488RDC) where it is reflected towards the scanning unit. Here, two scanning mirrors SMY1 and SMY2 (Cambridge Technologies, CT6800HPL with CTI CB6580 driver) achieve the scan along the *y*-axis while keeping the pupil plane stable at the position of the resonant *x*-scan mirror SMX (EOPC, SC-30, resonant optical scanner, 15 kHz, USA). Another beam expander consisting of the tube lens ($f_{TL}$=400mm) and an achromatic doublet ($f_3$=60mm) provides a slight over-illumination of the back focal plane of the objective (Carl Zeiss, Plan-Apochromat 63x/0.7 Oil). On the detection side, the returning light is descanned using scan mirrors SMX, SMY1 and SMY2. Fluorescent and back-scattered illumination light are separated by the dichromatic beam splitter (BS1) and the fluorescent light is expanded by lenses $L_4$ and $L_5$ – this is the additional intermediate beam expansion. To achieve confocal sectioning a pinhole can be placed between these lenses, as this is a conjugate plane of the sample plane. Since the ideal intermediate beam expansion depends on the Stokes shift of the imaged fluorophores, the magnification can be adjusted by choosing the focal lengths of the lenses $L_4$ and $L_5$. After this intermediate magnifying step, the emission light is guided to the same scanning unit to be rescanned. Lens $L_6$ finally directs the emission light to a camera (Andor Technology Inc., Neo sCMOS, Belfast) where a super-resolved image is captured by integrating over a full scan process.

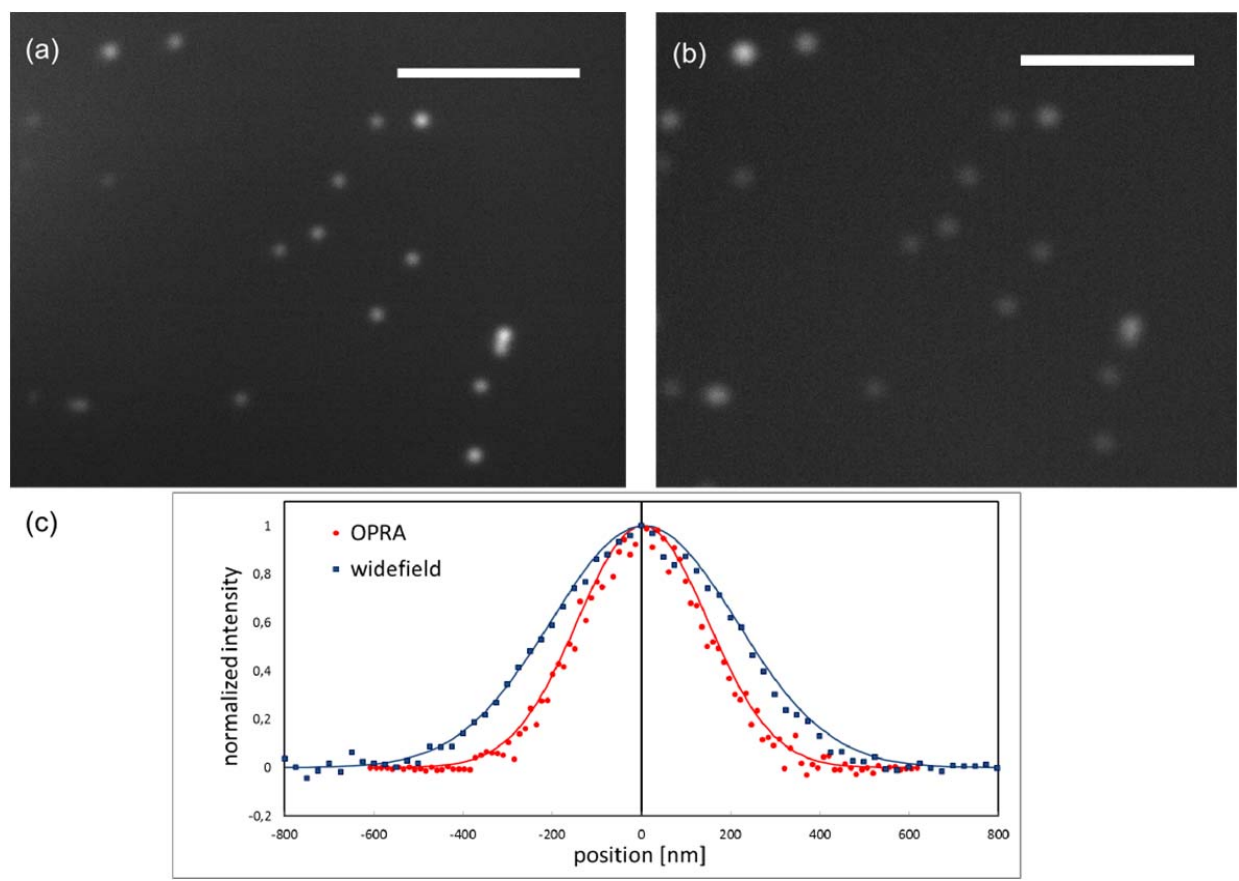

**Figure 3** Comparison of fluorescent beads imaged in OPRA and widefield (with an arrested scan system) mode. For determination of the FWHM, 8 beads were averaged and fitted with a 2D Gaussian function (solid lines). It is clearly visible that the FWHM of the scanned bead images is reduced. The FWHMs of the beads in the respective images was determined to amount to 489 ± 18 nm (widefield) and 326 ± 4 nm (OPRA). Scale bar 5µm.

RESULTS

To demonstrate the OPRA principle, fluorescent coated beads (FluoSpheres® Carboxylate-Modified Microspheres, 0.2 µm, Yellow-Green Fluorescent (505/515)) with a diameter of 200nm were imaged. For comparison a widefield excitation lamp (EXFO photonics solutions Inc., X-Cite series 120 Q) was coupled into the setup with an optional dichromatic beam splitter (Carl Zeiss Microscopy GmbH, FT 510), such that the same sample position could be imaged with both methods as shown in Fig. 3. For comparison 8 bead-images were analysed and the full-widths at half maxima (FWHM) of fitted 2D Gaussian functions were determined. The mean FWHM of measured beads in the widefield image (Fig. 3b) is determined to be (489± 18)nm and in the OPRA-image (326± 4)nm (Fig. 3a). Accounting for the 200nm diameter of the beads, theory predicts a FWHM of 297nm for the bead images acquired with OPRA, and 428nm for the widefield case. Thus theory and experiment agree to within about 10%. for the OPRA mode and 14% for the widefield detection. As the OPRA image theoretically contains the same number of photons as the widefield image but distributed to a sharper image, it is expected that the OPRA image also looks significantly brighter that the corresponding widefield (or confocal) image. However, since we used a separate illumination source to generate the widefield image, we could not do an appropriate comparison in this study.

DISCUSSION

In the presented paper a new method in fluorescence microscopy was introduced - OPRA microscopy. It realises super-resolved images with high detection efficiency. Compared to classical photon reassignment no post-processing is required as the summation and photon reassignment is a system inherent property of OPRA. This prevents artefacts (e.g. pixilation artefacts, additional read noise) and is even insensitive to small variations in the scanning process, as those will mostly lead to small brightness changes in the resulting image. This makes it useful for very fast imaging with the scanning-speed and the camera frame-rate as the only limiting factors. We showed that the principle improves the resolution in comparison to classical widefield microscopy and we derived the basic theory for OPRA performance. It should be noted that the required OPRA properties are also achievable with realisation methods other than intermediate magnification, such as the use of separate scan-units for illumination and detection light running at different speeds. OPRA can be adapted to various ratios of the sizes of excitation and emission PSF. Therefore the OPRA principle can also be used to realise versions of super-resolution modes like STED and RESOLFT [7]. For example a multi-spot STED or RESOLFT microscope is feasible with OPRA, where the super-resolved image is built up during the integration time of a single frame. Also the camera can be replaced by the human eye realising direct-view versions of STED or RESOLFT microscopy. Due to the simplicity and flexibility of the realisations, OPRA can enhance the performance of nearly every laser scanning microscope.

Even though OPRA improves photon reassignment by its all-optical realisation, it should be noted that a full pinhole plane scan dataset of images in dependency of scan and image coordinate is richer and allows for better ways of image processing. This ranges from the ability retroactively to select the pinhole size to optimised strategies such as weighted averaging in Fourier space and combined deconvolution [8].

REFERENCES

1. C. J. R. Sheppard, "Super-resolution in confocal imaging", Optik (Stuttg.) 80, 5354 (1988).

2. James, B., Pawley, (Eds.), Handbook of Biological Confocal Microscopy. Plenum Press, New York (1990).

3. M. Minsky, "Microscopy apparatus", U.S. patentt 3,013,467 (1961).

4. I. J. Cox, C. J. R. Sheppard, and T. Wilson, "Improvement in resolution by nearly confocal microscopy", Appl.Opt. 21(5), 778781 (1982).

5. C. B. Müller and J. Enderlein, "Image Scanning Microscopy", Physical Review Letters 104(19), 14 (2010)

6. A. G. York, S. H. Parekh, D. D. Nogare, R. S. Fischer, K. Temprine, M. Mione, A. B. Chitnis, C. A. Combs, and H.Shroff, "Resolution doubling in live, multicellular organisms via multifocal structured illumination microscopy",Nat. Methods 9(7), 749754 (2012).

7. S. Hell, "Toward fluorescence nanoscopy", Nature Biotechnology 21, (1347 – 1355) (2003).

8. R. Heintzmann, V. Sarafis, P. Munroe, J. Nailon, Q. S. Hanley and T. M. Jovin, "Resolution enhancement by subtraction of confocal signals taken at different pinhole sizes", Micron 34(6-7), 293300 (2003).